# Temperature dependent creation of nitrogen-vacancy centers in CVD diamond layers

A. Tallaire[1,a], M. Lesik[2], V. Jacques[2,b], S. Pezzagna[3], V. Mille[1], O. Brinza[1], J. Meijer[3], B. Abel[4], JF. Roch[2], A. Gicquel[1] and J. Achard[1]

[1] *Laboratoire des Sciences des Procédés et des Matériaux (LSPM), Université Paris 13, Sorbonne Paris Cité, CNRS, 93430 Villetaneuse, France*

[2] *Laboratoire Aimé Cotton, CNRS, Université Paris-Sud and ENS Cachan, 91405 Orsay, France*

[3] *Department of Nuclear Solid-State Physics, University Leipzig, 04103 Leipzig, Germany*

[4] *Leibniz-Institute of Surface Modification (IOM), Chemical Department, 04318 Leipzig, Germany*

In this work, we explore the ability of plasma assisted chemical vapor deposition (PACVD) operating under high power densities to produce thin high-quality diamond layers with a controlled doping with negatively-charged nitrogen-vacancy (NV$^-$) centers. This luminescent defect possesses specific physical characteristics that make it suitable as an addressable solid-state electron spin for measuring magnetic fields with unprecedented sensitivity. To this aim, a relatively large number of NV$^-$ centers ($> 10^{12}$ cm$^{-3}$) should ideally be located in a thin diamond layer (a few tens of nm) close to the surface which is particularly challenging to achieve with the PACVD technique. Here we show that intentional temperature variations can be exploited to tune NV$^-$ creation efficiency during growth, allowing engineering complex stacking structures with a variable doping. Because such a temperature variation can be performed quickly and without any change of the gas phase composition, thin layers can be grown. Measurements show that despite the temperature variations, the luminescent centers incorporated using this technique exhibit spin coherence properties similar to those reached in ultra-pure bulk crystals, which suggests that they could be successfully employed in magnetometry applications.

a) **Electronic mail:** alexandre.tallaire@lspm.cnrs.fr
b) **Electronic mail:** vjacques@ens-cachan.fr

Keywords: NV centers; single crystal diamond; nitrogen doping; crystal growth; Plasma assisted chemical vapour deposition

## 1. Introduction

The negatively-charged nitrogen-vacancy (NV⁻) defect in diamond has outstanding optical and electron spin properties, which make it a promising system for a broad range of applications including quantum information processing [1], and highly sensitive magnetometry [2,3]. These applications mainly rely on the long coherence time of the NV⁻ defect electronic spin, which is limited by magnetic interactions with surrounding paramagnetic impurities in the diamond lattice [4]. The recent progresses of diamond growth using Plasma Assisted Chemical Vapor Deposition (PACVD) techniques now allows to obtain extremely high-purity diamond samples [5,6], leading to coherence time exceeding milliseconds under ambient conditions [7]. Besides these purity requirements, for most applications a substantial degree of spatial localization of NV⁻ defects in the diamond host material is necessary. For instance, in wide-field imaging magnetometry, nanometer-thin layers highly doped with NV⁻ defects should be located near the surface [8,9]. These stringent requirements are setting an increasing pressure to the diamond synthesis capabilities and diamond technologies.

There are currently two approaches to introduce nitrogen atoms into diamond in order to generate NV⁻ defects. *Ex-situ* ion-beam nitrogen implantation followed by annealing allows for a spatial control of the nitrogen atom's position ultimately limited by ion channeling and straggling effects [10]. For nitrogen atoms implanted near the diamond surface with a few keV energy, the depth resolution is typically in the range of few nanometers [8,11]. However, unwanted paramagnetic defects presumably created during the implantation process reduce the spin coherence time of implanted NV⁻ defects [11], although significant improvements can be achieved with optimized irradiation and annealing procedures [12,13]. On the other hand, NV⁻ centers can be directly grown-in by intentionally adding nitrogen to the gas phase during diamond growth by PACVD [5,6]. Using this method, there is no residual radiation damage and thus no need to rely on post-implantation procedures. The main drawback though is the limited degree of depth control of the created NV⁻ defects in the crystal. Indeed even after turning off the $N_2$ input flow, nitrogen remains inside the plasma reactor for a fairly long period of time and keeps incorporating into the crystal. Consequently, thin diamond layers with a high content of NV⁻ defects cannot be easily achieved. This issue is somewhat similar to that reported for delta boron-doped diamond films [14], although in this latter case, obtaining thin, sharp and heavily doped stacks is further hampered by strong memory effects of the boron impurity in the reactor.

There have been only a few studies reporting the CVD growth of thin N-doped layers with thicknesses ranging from a few nanometers [15,16] up to 100 nm [17]. Ishikawa et al. have shown that NV⁻ centers placed in a 100 nm-thick isotopically purified layer exhibit ms-long coherence time together with a 10-fold improvement in line-width as compared with implanted centers [17]. When the thickness of the layer is further decreased (5 nm), the proximity of the surface has a strong influence on the coherence time which is reduced to a few tens of μs only [16]. Higher coherence times of several hundreds of μs were reached for an electron irradiated and annealed structure consisting of a 2 nm-thick $^{15}$N-doped layer sandwiched between high-purity $^{12}$C diamond layers [15]. These growth techniques with a nanometer-scale resolution have already shown promising results in magnetometry. They rely on extremely low growth rates (a few nm/h) thanks to the use of low power densities (typically 750W, 30 mbar) together with low methane additions (< 0.5 %). The main drawback though is that at such a low power, $N_2$ dissociation is very limited and cannot lead to a high NV⁻ concentration even when large amounts of nitrogen are introduced in the gas phase (several %). Since the magnetic field sensitivity scales as $1/\sqrt{N}$, where $N$ is the number of sensing spins involved in the measurement [18], a strong improvement is expected if one could incorporate large ensembles of NVs in thin layers. In Ref. [15-17], NV⁻ concentrations thus only were in the range $10^8$ to $10^{12}$ cm$^{-3}$. At low-power densities it is also difficult to obtain a high crystalline quality especially if one wants to grow a thick buffer layer (> 10 μm) [19,20]. Therefore growth of the active layers must be performed on expensive type *IIa* electronic grade material.

The main purpose of this work is thus to assess the ability of the high-power PACVD technique to control the amount and depth distribution of NV⁻ centers which can ensure high crystalline quality layers together with a good NV⁻ creation efficiency. To this end, we propose a new strategy based on the temperature dependence of nitrogen incorporation.

## 2. Experimental details

All the samples studied in this work were grown on low-cost type *Ib* HPHT substrates purchased from *Sumitomo*. In order to ensure that there is no parasitic background luminescence from the substrate, a thick buffer layer (> 40 μm) was grown using our optimal deposition conditions, as reported elsewhere [20]. They typically involve high-power densities (3.5 kW microwave power at a pressure of 250 mbar), a temperature of 850 °C and no intentional addition of $N_2$. These growth conditions result in high-quality material with no detectable NV⁻ centers. Active layers containing NV⁻ centers were then directly overgrown onto this buffer layer in the same reactor, without any growth interruption as interruptions are known to promote the incorporation of unwanted impurities and extended defects [21].

For sample A, a 3-layer structure consisting of two nitrogen-doped layers with 50 and 100 ppm of intentionally added $N_2$ into the gas phase, separated by a nitrogen-free layer was grown as shown in Fig. 1a. The arrows on the figure illustrate that growth occurs both vertically and laterally from the substrate. This sample aims at evaluating the dependence of NV⁻ creation on the amount of added $N_2$ to the gas phase. Obtaining a sharp transition between stacked layers when changing the gas phase composition is indeed hampered by long residence times $t_r$ of gas species in the plasma chamber as defined by:

$$t_r = \frac{V_{ch} \times P}{F_{tot} \times P_{atm}}$$

where $V_{ch}$ is the plasma chamber volume, $P$ and $P_{atm}$ are the operating and atmospheric pressures respectively, and $F_{tot}$ is the total gas flow injected in the chamber. In our system, a total gas flow of 500 sccm was injected in a reactor volume of about 22 liters, which corresponds to a residence time $t_r \approx 11$ minutes. On the basis of this estimation, the growth duration of each doped layer in sample A was set to 30 minutes so that it gives enough time for the gas mixture to reach a steady-state.

NV⁻ creation efficiency was then varied at a constant gas phase composition by changing the growth temperature in a second diamond structure labeled B (Fig. 1b). It consisted of an 11-layer stack grown using an increasingly low growth temperature (30 min in the range 780-880 °C) separated by high temperature spacers (45 min at 905 °C). The temperature variation was achieved by quickly dropping or raising the reactor pressure in the 220-250 mbar range. A change of a few mbar strongly affects gas temperatures in the plasma [22] which in turn directly translates into a change in surface temperature since the sample is placed on a cooled stage in this high-power PACVD configuration. For sample B, no nitrogen was intentionally added to the gas phase but it is known from previous experiments that our set-up contains a low level of background $N_2$ impurities (< 1 ppm). Electron irradiation ($5 \times 10^{15}$ cm⁻² at 10 MeV) followed by annealing for 1 h at 800 °C was then carried out in order to enhance the appearance of NV⁻ luminescent defects by the creation and diffusion of vacancies [23].

In order to assess if the temperature dependency technique can be used to obtain thin layers highly doped with NV⁻ centers, another structure labeled C (Fig. 1c) was prepared with addition of 2 ppm of $N_2$. It included low temperatures layers grown at 760 °C with an increasingly short deposition time (from 2 h down to 5 min) and spaced by high temperature layers grown at 840 °C. In this case, because the input gas flow remained constant, there was no issue arising from long residence times allowing short growth to be performed. No irradiation treatment was performed for this sample.

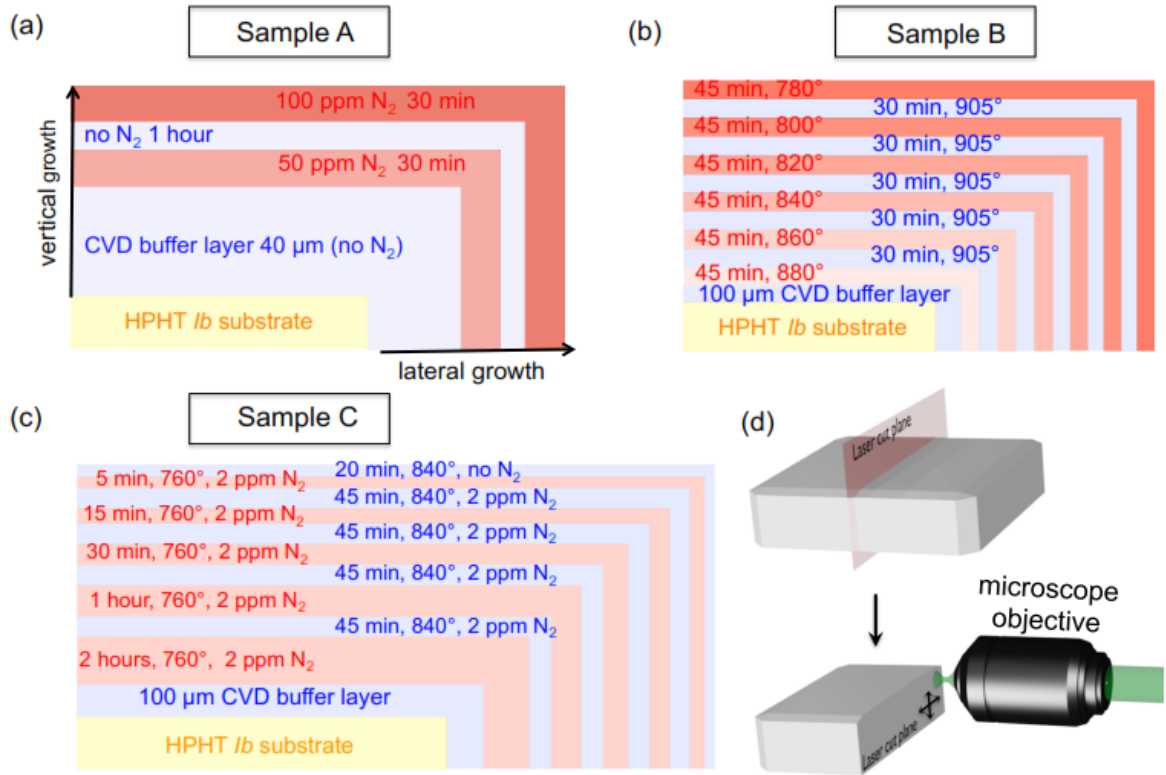

*Fig. 1*: Sketch of the cross-sections of different structures analyzed. (a) CVD layers with a variable amount of added $N_2$ (sample A), (b) CVD layers grown by varying the growth temperature (sample B) and (c) CVD layers grown at different temperatures and growth times. In (b) and (c), the blue and red layers correspond to high and low growth temperatures respectively. The black arrows indicate the vertical and lateral growth directions from the substrate. (d) Schematics illustrating how the cross-section of the sample is prepared by laser-cutting. The PL response is analyzed using a scanning confocal microscope in the vertical and lateral growth sectors (by exploring along the vertical and horizontal double arrows).

The surface morphology of the films was first observed using 3D laser microscopy (*Keyence VK9700*). A cross-section was then obtained by laser-cutting and polishing the samples. The photoluminescence (PL) properties of the cross-section were studied with a customized confocal microscope under optical excitation at 532 nm, as shown in Fig. 1d. Details about the experimental set-up can be found in Ref. [24]. This geometry enables analyzing the profile of NV-doped layers with a spatial resolution limited by the lateral resolution of the confocal microscope (~ 300 nm). To avoid any parasitic influence from the surface, the excitation laser was focused a few micrometers below the cross-section's surface.

### 3. Results and discussion

3.1. Control of $NV^-$ creation by varying the added $N_2$ amount

The surface morphology of sample A (Fig. 2a) shows a high density of macro-steps which is commonly observed in the presence of a high amount of nitrogen [25]. PL images of the cross-section recorded in the vertical and lateral growth sectors are shown in Figs. 2b and 2c. As expected, a high PL signal is observed in the nitrogen-doped layers. In these regions, the PL spectrum exhibits a broadband emission with a characteristic zero-phonon line (ZPL) at 637 nm, which is the signature of the negatively-charged $NV^-$ defect (Fig.2d).

Along the vertical growth sector, preferential incorporation of NVs at step edges [26] occurred leading to the formation of inclined stripes which is consistent with the step-bunching observed. In the lateral growth sector, $NV^-$ creation efficiency was on average higher if one excludes the stripes of the vertical growth sector where the intensity of the $NV^-$ luminescence was very high, and the distribution was more uniform. Such a discrepancy might be related to a non-uniform temperature distribution across the sample although this would need further investigation.

The resulting density of $NV^-$ defects was roughly inferred by comparing PL intensity to that measured from a single $NV^-$ defect, leading to 60 and 130 $NV^-$ in about 1 μm$^3$ for the layers doped with 50 and 100 ppm of $N_2$ in the gas phase respectively (Fig. 2e). This corresponds to a creation yield of about $10^{-5}$ (i.e. the concentration of $NV^-$ to that of $N_2$ in the gas phase) which is consistent with our previous report [5]. As expected, the growth rate is strongly enhanced by the presence of $N_2$ with values of 5, 18 and 25 μm/h for doping with 0, 50 and 100 ppm respectively.

The analysis of sample A highlights that under high plasma power densities it is possible to achieve a tunable and reasonably high $NV^-$ density ($> 10^{12}$ cm$^{-3}$) even for a low $N_2$ addition. However, reducing layer thickness and improving sharpness remains difficult due to the high growth rates and long residence times of gas species. In order to circumvent this problem, we propose that rather than changing the gas phase mixture, the growth conditions could be quickly adjusted in order to vary the NV⁻ creation efficiency [27].

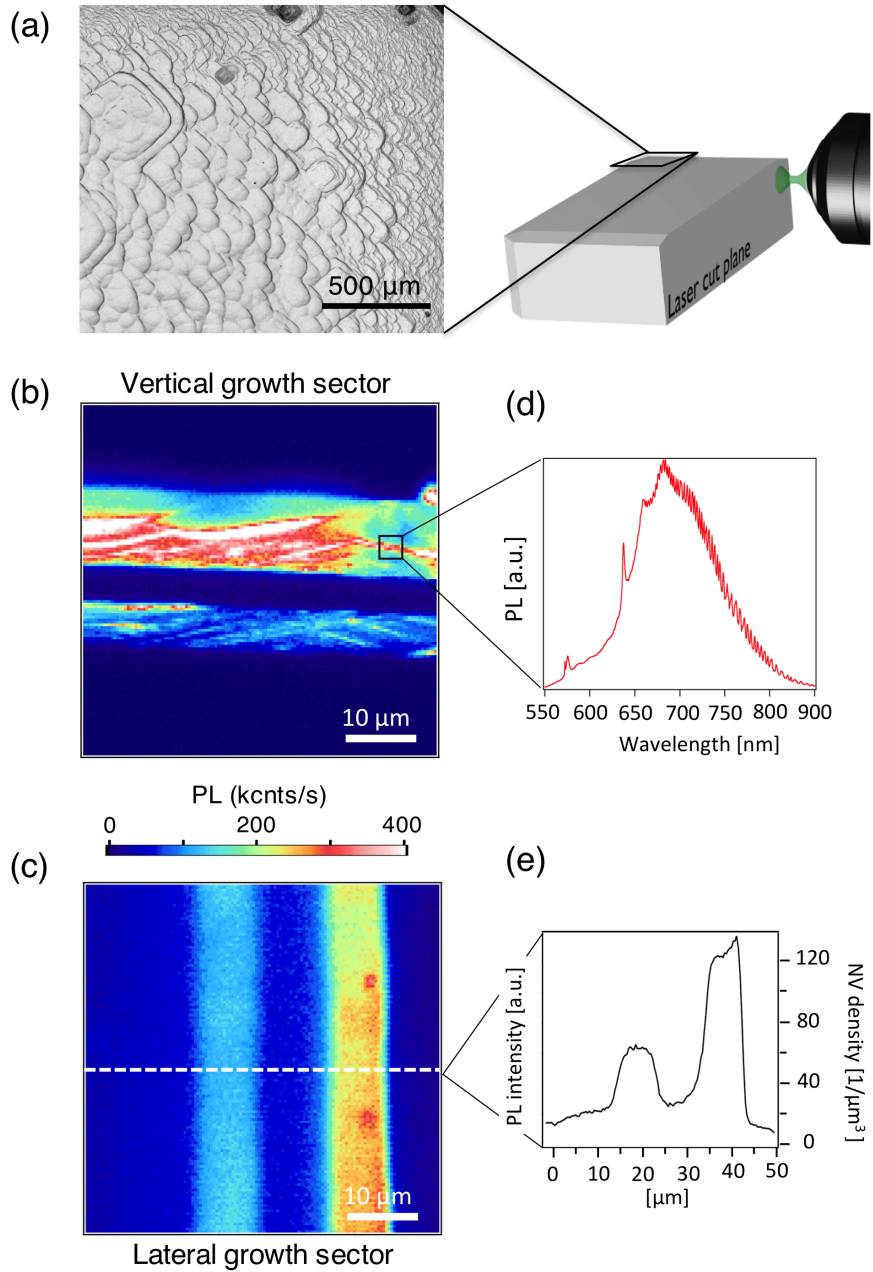

*Fig. 2.* (a) Laser microscope image of the surface morphology of sample A grown with intentional addition of $N_2$. The PL response of sample A's cross section is analyzed with a scanning confocal microscope. PL raster scans recorded in the vertical (b) and lateral (c) growth sectors. (d) PL spectrum recorded in the 100 ppm nitrogen-doped layer, mainly corresponding to the emission from negatively-charged NV⁻ defects with a ZPL at 637 nm. We note a residual emission from the neutral NV° defect, with zero-phonon line at 575 nm. (e) PL intensity linecut along the white dashed line shown in (c).

### 3.2. Effect of growth temperature on the creation of nitrogen related defects

Layers grown with increasingly low growth temperatures were stacked in sample B and spaced by high-temperature layers. Because no $N_2$ was intentionally added, the surface morphology of the film remained very smooth with no evidence of step bunching (Fig. 3a). PL analysis performed in the cross-section, along the vertical growth sector did not reveal any luminescent $NV^-$ center. In order to enhance their appearance, electron irradiation and annealing were thus performed in order to create vacancies that will later diffuse and recombine with incorporated substitutional nitrogen atoms originating from background impurities.

From the PL raster scan of Fig. 3b and the averaged PL intensity of Fig. 3c, it appears that the lower the growth temperature, the higher the $NV^-$ luminescence gets, with a roughly proportional dependence. The limited difference in temperature between successive layers (20 °C only) may however sometimes be locally masked by other effects such as surface roughening. For example L3 and L4 exhibit almost similar luminescence intensity in the analyzed region. Comparison between the vertical and lateral growth sectors showed that, as for the previous sample, $NV^-$ emission was more pronounced in the latter.

This result shows that it is possible to tune $NV^-$ incorporation efficiency by only a slight temperature variation. The luminescence in L6 was almost 3 times higher than that in the high-temperature spacer. The contrast could be further improved by increasing the temperature range to 700-1000 °C in which reasonably good diamond crystal quality can be obtained. Since most luminescent centers observed in this sample were created by the post-treatment, it is believed that under low growth temperature, the total amount of nitrogen incorporated is increased in the form of NVs but also $N_s$ defects. For the purpose of magnetometry, a much higher density of $NV^-$ centers could be obtained by intentional addition of several ppm of $N_2$ during growth. In that case, irradiation post-treatment might not even be necessary to enhance $NV^-$ appearance.

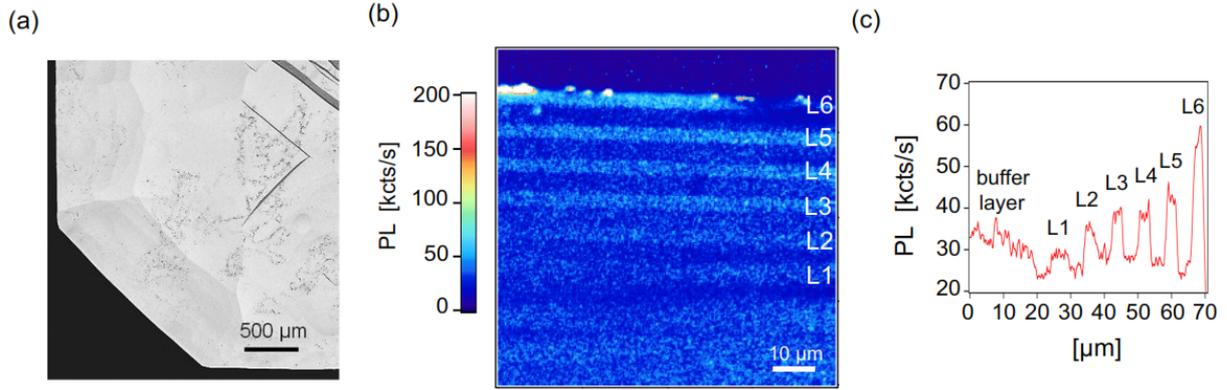

***Fig. 3.*** *(a) Laser microscope image of the top surface morphology of sample B; (b) PL raster scan of the cross-section acquired along the vertical growth sector and allowing identifying the low-temperature layers labeled L1 to L6; (c) Averaged PL intensity.*

3.3. Growth of thin NV-containing layers by varying the growth temperature

The main advantage of varying the growth temperature to tune $NV^-$ incorporation efficiency is that, to the contrary of a gas phase composition variation, it can be done quickly allowing reducing the thickness and sharpness of the doped layers. To this aim, another structure (sample C) was grown using increasingly low growth durations for the low-temperature layers and intentional addition of 2 ppm of $N_2$. This low nitrogen amount was not sufficient to generate a detectable density of $NV^-$ centers in the vertical growth sector, and irradiation/annealing post-treatment could not be carried out. To evaluate the distribution of $NV^-$ centers, analysis was thus performed on the lateral growth sector where incorporation efficiency is much higher, as previously observed.

The surface morphology (Fig. 4a) shows only limited step-bunching. In the PL raster scan of Fig. 4b the low-temperature layers labeled L1 to L5 are visible. In the considered range (840-760 °C) a $NV^-$ density variation by a factor of ≈ 4, comparable to that of the previous sample was found. Growth rates were estimated to be around 7 and 10 µm/h for the low and high temperature layers respectively. The layer grown with the shortest deposition time (5 minutes only) exhibits a PL profile with a linewidth (FWHM) in the order of ≈ 850 nm (inset of Fig. 4b and linecut of Fig. 4c). Considering the diffraction limited resolution of the confocal microscope, thickness of the NV-doped layer is therefore in the range of ≈ 500 nm. Although this is higher than the thickness reported in ref. [15-17], it is believed that further decreasing both the growth rate and the deposition time could lead to much thinner layers with the advantage of keeping a high $NV^-$ density (> $10^{12}$ NVs/cm$^3$). This experiment can therefore be regarded as a proof-of-principle that the temperature dependence of $NV^-$ creation can be exploited for engineering thin nitrogen-doped layers in high-purity material.

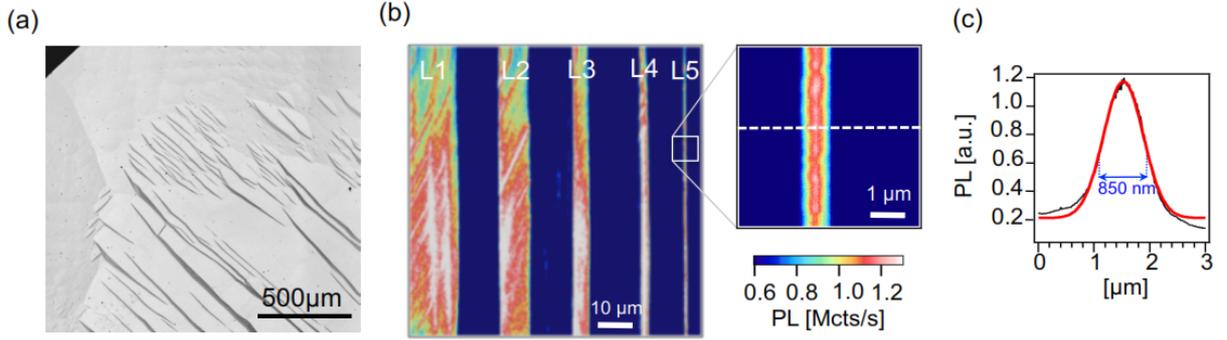

***Fig. 4.*** *(a) Laser microscope image of the top surface morphology of sample C; (b) PL raster scan of the cross-section acquired in the lateral growth sector and showing the presence of 5 low-temperature layers labeled L1 to L5; The inset shows a zoom into the thinnest NV-doped layer. (c) PL intensity linecut along the white dashed line shown in (b). The solid line is data fitting with a Gaussian function, leading to a full-width at half maximum of ≈ 850 nm.*

3.4. Spin properties of NV⁻ centers created in low-temperature layers

It is relevant to estimate if NV⁻ defects incorporated in low temperature layers exhibit spin properties comparable to those reached in high-quality bulk diamond samples. To this end, electron spin resonance (ESR) measurements were performed at room temperature by monitoring the spin dependent PL of the NV⁻ defect while applying a radiofrequency field [28]. A typical ESR spectrum recorded at zero magnetic field from a NV-doped layer of sample C is shown in Fig. 5a (top panel). A characteristic resonance at ≈ 2.87 GHz is clearly observed, corresponding to the zero-field splitting between $m_s = 0$ and $m_s = \pm1$ spin sublevels of the NV⁻ defect [29]. Applying a static magnetic field along the [111] axis of the diamond crystal enables (i) lifting the degeneracy of $m_s = \pm1$ spin sublevels and (ii) dividing the NV⁻ defect ensemble into two sub-groups of crystallographic orientations which experience different Zeeman splitting (Fig. 5a, lower panel). A high resolution zoom into one of the ESR transition clearly reveals the hyperfine structure associated with the ¹⁴N nucleus of the NV⁻ defect (inset of Fig. 5a). We now focus on the ESR line with lower frequency corresponding to the ESR transition for the subset of [111]-oriented NV⁻ defects. The spin coherence time of this subset of NV⁻ defects was measured by applying a Hahn echo sequence. As shown in Fig. 5b, the spin-echo signal exhibits characteristic collapses and revivals of the electron spin coherence induced by Larmor precession of ¹³C nuclear spin impurities [30]. The decay of the envelope indicates a coherence time $T_2 = 232 \pm 7$ μs, which is comparable to reference values in high-quality diamond samples with the same natural abundance of ¹³C (1.1%) [31]. Spin coherence properties are therefore preserved for thin NV-doped layers obtained through temperature variations during CVD growth.

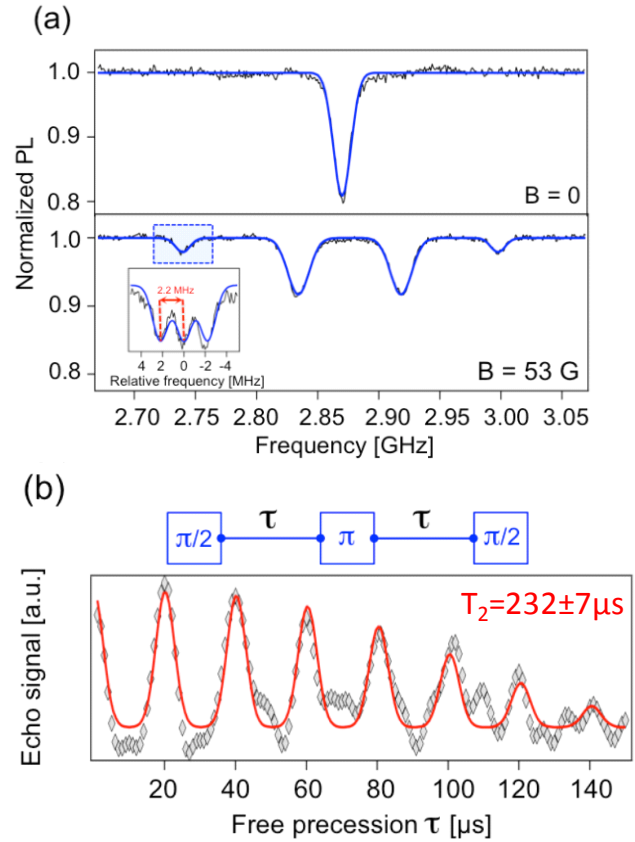

*Fig. 5.* (a) ESR spectra recorded from the thinnest NV-doped layer (labeled L5 in Fig. 4b) at zero field (top panel) and with a static magnetic field B = 53 G applied along the [111] axis of the diamond crystal (lower panel). The inset shows a high resolution zoom into one of the ESR lines revealing the characteristic hyperfine structure associated to the $^{14}N$ nuclear spin. (b) Spin echo signal recorded for the subset of [111]-oriented NV⁻ defects with a magnetic field B = 53 G applied along the [111] axis. The spin echo sequence [π/2 − τ − π − τ − π/2] consists of resonant radiofrequency π/2 and π pulses separated by a variable free evolution duration τ. The solid line is data fitting with a sum of Gaussian peaks modulated by a decay envelope [29].

### 4. Conclusion

In summary, the ability of the PACVD technique operating under high power densities to produce thin NV-doped diamond layers was assessed. One of the main advantages is that a reasonably high NV⁻ density (> $10^{12}$ cm$^{-3}$) can be reached for moderate additions of $N_2$ in the gas phase (in the ppm range) while keeping a good crystalline quality. However the long residence times and high growth rates involved in this approach do not allow for an accurate depth control of NV⁻ distribution. Instead, it is proposed that the temperature dependency of nitrogen incorporation can be usefully exploited in order to finely tune nitrogen doping at a constant $N_2$ concentration in the gas phase. Because such a change can be quickly achieved, complex stacking sequences with tuned $NV^-$ densities can be engineered. Despite being grown-in under different temperatures, $NV^-$ centers keep their excellent coherence properties comparable to a more conventional process. Layers obtained by this technique will find useful applications in magnetic field sensing and imaging. Further improvements will include growth of isotopically pure NV-doped diamond layer with a further reduced thickness, combined with recently reported deterministic control of the $NV^-$ defect orientation in the crystal lattice [32,33,34].

**Acknowledgements**

We thank J. P. Tetienne and W. Knolle for fruitful discussions and experimental assistance. This research has been partially funded by the European Community's Seventh Framework Programme (FP7/2007-2013) under Grant Agreement n◦ 611143 (Diadems) and by the French Agence Nationale de la Recherche through the project Advice (ANR-2011-BS04-021-03).